\documentclass{ws-p8-50x6-00}

\begin{document}

\title{Annihilations from the Galactic Centre}

\author{G.Bertone and J.Silk}

\address{Department of Astrophysics, University of Oxford, \\NAPL
Keble Road, Oxford OX13RH, United Kingdom\\E-mail: bertone@astro.ox.ac.uk,
silk@astro.ox.ac.uk}

\author{G.Sigl}

\address{ Institut d'Astrophysique, F-75014 Paris, France\\
E-mail: sigl@iap.fr}  

\maketitle

\abstracts{A massive black hole is present at the centre of our galaxy and
inevitably accretes dark matter particles, creating a region of very high
particle density. The annihilation rate  is enhanced  with a large number
of $e^+e^-$ pairs produced either directly or by sucessive decays of
mesons. We evaluate the synchrotron emission (and self-absorption)
associated with
 the propagation of these particles through the galactic magnetic field
and
  give constraints on the values of mass and cross section of
the  dark matter particles.}

\section{Introduction}

There is  convincing evidence for the existence of an unseen non-baryonic
component in the energy density of the universe. The most promising dark
matter candidate appear to be weakly interacting massive particles (WIMPs)
and in particular the so-called neutralinos, arising in supersymmetric
scenarios  (for a review of particle candidates for dark matter see e.g.
\cite{ellis}). The annihilation of these X-particles would produce quarks,
leptons, gauge and Higgs bosons and gluons. In particular
 $e^+e^-$ pairs are produced either directly or by successive decays of
mesons, and lose their energy through synchrotron radiation as they
propagate in the galactic magnetic field. This radiation is expected to be
greatly enhanced in the proximity of the galactic centre, where the
existence of a massive black hole creates a region of very high dark
matter particle density and consequently a substantial increase in the
annihilation rate and 
in the ensuing synchrotron radiation.

\begin{figure}[t]
\epsfxsize=27pc
\epsfbox{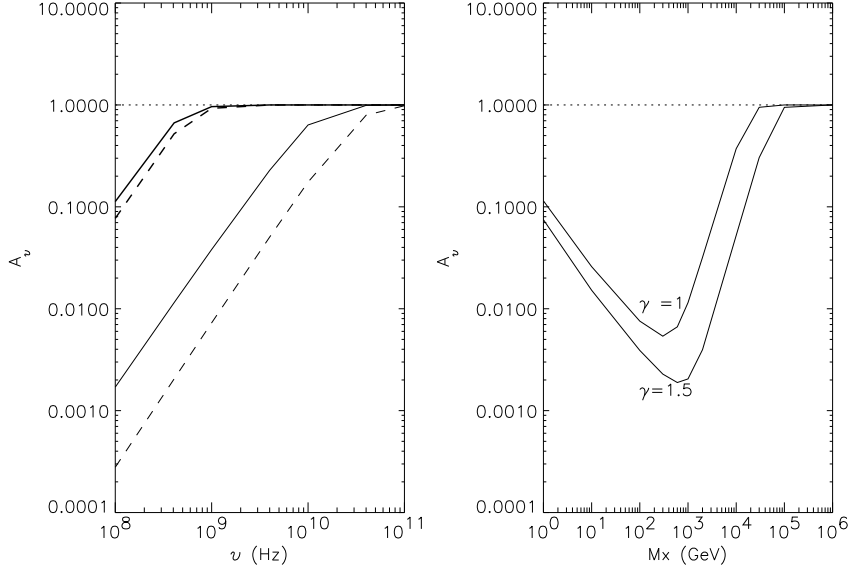} 
\caption{Left panel: $A_{\nu}$ as a function of frequency for $m_X=1$TeV.
The two upper curves correspond to the cross section $\sigma v=10^{-28}/m_X^2(\rm GeV)\,\rm cm^3s^{-1}$,
close to the unitarity limit; the two lower curves correspond to
$\sigma v=10^{-38}/m_X^2(\rm GeV)\,\rm cm^3s^{-1},$ a cross section more typical for wimps.
Results for two values of the density profile are shown in each case:
$\gamma=1$, for solid curves and $\gamma=1.5$ for dashed ones.
Right panel: $A_{\nu}$ as a function of
the particle mass for $\nu$=408MHz, $\sigma v=10^{-38}/m_X^2(\rm GeV)\,\rm cm^3s^{-1}$
and two
values of $\gamma$. \label{freq}}
\end{figure}
\section{Dark matter distribution around the Galactic Centre}

There is strong evidence for the existence of a massive compact object
lying within the inner $0.015 \rm  pc$ of the galactic centre (see \cite{yusef}
and references therein). This object is a compelling candidate for a
massive black hole, with  mass $M=2.6 \pm 0.2 M_{\odot}$.
The galactic halo density profile is modified in the neighbourhood of the
galactic centre by the adiabatic process of accretion towards the central
black hole. If we consider an initially power-law type profile of index
$\gamma$, as predicted by high resolution N-body simulations \cite{nava},
the corresponding dark matter profile after accretion is modified to 
\begin{equation}\rho' = \left[ \alpha_{\gamma} \left( \frac{M}{\rho_D
D^3}\right)^{3-\gamma} \right]^{\gamma_{sp}-\gamma} \rho_D \; g(r) \left(
\frac{D}{r} \right)^{\gamma_{sp}}
\label{modaccre}\end{equation}
where $\gamma_{sp}=(9-2\gamma)/(4-\gamma)$, $D$ is the solar distance from
the galactic centre and $\rho_D=0.24 GeV/c^2/cm^3$ is the corresponding
density. The factors $\alpha_\gamma$ and $g_\gamma(r)$ cannot be
determined analytically (for approximate expressions and numerical values
see \cite{gondolo}). Expression  \ref{modaccre} is only valid in a
central region of size $R_{sp}=\alpha_\gamma D (M/\rho_D
D^3)^{1/(3-\gamma)}$ where the central black hole dominates the
gravitational potential.
\begin{figure}[t]
\epsfxsize=27pc 
\epsfbox{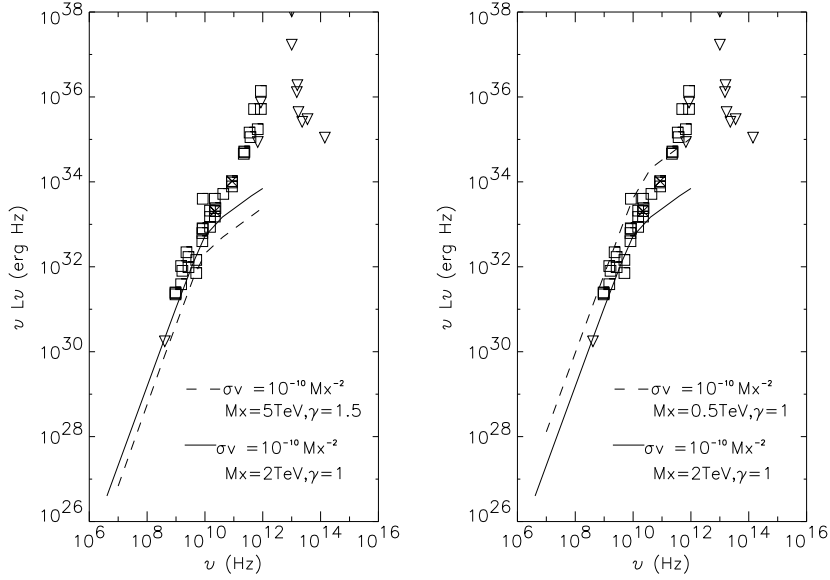}
\caption{Comparison of Sgr A* (see Narayan {\em et al}(1998)) observed
spectrum with expected fluxes. The values of particle mass and cross
section were choosen to fit the experimental data normalisation.
\label{predi1}}
\end{figure}
If we take into account the annihilation of dark matter particles, the
density cannot grow to arbitrarily high values, the maximum density being
set
by  the value 
\begin{equation}
\rho_{core}=m/\sigma v t_{BH}
\label{roco}
\end{equation}
where $t_{BH}\approx 10^{10}\rm years$ is the age of the central black hole.
The final profile, resulting from the adiabatic accretion of annihilating
dark matter onto a massive black hole, is
\begin{equation}
\rho_{dm}(r)= \frac{\rho'(r) \rho_{core}}{\rho'(r)+\rho_{core}}
\end{equation}
following a power-law profile for large values of r, and with a flat core
of density $\rho_{core}$ and  dimension
 \begin{equation}
R_{core}=R_{sp} \left( \frac{\rho(R_{sp})}{ \rho_{core}} \right)
^{(1/\gamma_{sp})}
\end{equation}

\section{Constraints from Synchrotron Emission}
Among the products of annihilation of dark matter particles, there are
energetic electrons and positrons, which are expected to produce
synchrotron radiation in the magnetic field around the galactic centre.
The pair production spectrum is determined by the quark fragmentation
function and has
 been evaluated using the MLLA approximation (see \cite{mlla} for
details). 
The galactic magnetic field can be modeled by
making  the 'equipartition assumption', where the magnetic, kinetic and
gravitational energy of the matter accreting on the central black hole are
in approximate equipartition (see \cite{melia}). In this case, the
magnetic field can be expressed as
\begin{equation}
B(r) = 1\mu G \left( \frac{r}{\mbox{pc}} \right)^{-5/4}
\label{mag}
\end{equation}

 Most of the annihilations occur at very small distances from the centre,
typically at $\approx min(R_{core}, 10 R_s)$, i.e. in a region with
magnetic fields of the order of $> 1G$. Under these conditions, comparable
to the size of the region where most of the annihilations occur, the
electrons lose their energy almost in place.
To compute the synchrotron luminosity resulting from
the propagation of $e^{\pm}$ in the galactic magnetic field, we need to
evaluate their energy distribution in the magnetic field,
following~\cite{gondolo2}
\begin{equation}
 \frac{dn}{dE} = \frac{\Gamma Y_e(>E)}{P(E)} f_e(r)\label{dn}
\end{equation}
where $\Gamma$ is the annihilation rate
\begin{equation}
\Gamma = \frac{\sigma v}{m_x^2} \int_0^{\infty} \rho_{sp}^2 4 \pi r^2 \;\;
dr,
\end{equation}
the function $f_e(r)$ is given by
 \begin{equation}
f_e(r)= \frac{\rho_{sp}^2}{  \int_0^{\infty} \rho_{sp}^2 4 \pi r^2 \;\;
dr}
\end{equation}
and 
 \begin{equation}
P(E)= \frac{2 e^4B^2 E^2}{3 m_e^4 c^7}
\end{equation}
The quantity $Y_e(>E)$ is the number of $e^+e^-$ with energy above E
produced per annihilation, which depends on the annihilation modes. The
energy dependence of $Y_e(>E)$ can be neglected,
and we  estimate $Y_e(>E)$ by the number of charged particles produced in
quark fragmentation (see \cite{mlla}).
We obtain a simple analytical expression for 
the total synchrotron luminosity
\begin{equation}
L_{\nu} \approx \frac{9}{8} \left(\frac{1}{0.29 \pi} \frac{ m_e^3 c^5}{e}
\right)^{1/2} \frac{\Gamma Y_e(>E)}{\sqrt{\nu}} \; I
\label{lum}
\end{equation}
where 
\begin{equation}
I = \int_0^{\infty} dr \;\; 4 \pi r^2 f_e(r) B^{-1/2}(r) 
\end{equation}

 The synchrotron self-absorption coefficent is defined  \cite{Rybicki} by
\begin{equation}
A_{\nu}= \frac{1}{a_{\nu}} \int_{0}^\infty (1-e^{-\tau(b)}) 2 \pi b \;\;
db
\label{anu}
\end{equation}
where $\tau(b)$ is the optical depth as a function of the cylindrical
coordinate $b$
\begin{equation}
\tau(b)= a_{\nu} \int_{d(b)}^{\infty} f_e(b,z) \;\;dz
\label{la}
\end{equation}
and the coefficent $a_{\nu}$ is given by
\begin{equation}
 a_{\nu} =\frac{e^3 \Gamma B(r)}{9 m_e \nu^2} \int_{m_e}^{m} E^2 \frac
{d}{dE} \left[ \frac{Y_e(>E)}{E^2 P(E)} \right] F\left( \frac{\nu}{\nu_c}
\right) \;\; dE
\end{equation}
The final luminosity is obtained by multiplying eq.~(\ref{lum}) with
$A_{\nu}$ given by eq.~(\ref{anu}). It is evident that in the limit of
small optical depths the coefficent $A_{\nu} \rightarrow 1$, as can be
seen by expanding the exponential. We find for $a_{\nu}$ the following
expression
\begin{equation}
a_{\nu} =\frac{\Gamma Y}{4 \pi} \frac{c^2}{\nu^3}
\end{equation}
which can in turn be used to evaluate $\tau(b)$ in eq.~(\ref{la}) and
$A_{\nu}$ in eq.~(\ref{anu}).

We first evaluate the self-absorption coefficent for selected values of
the mass, as a function of frequency. The  coefficent  (see left part of
fig. \ref{freq}) grows from very low values and then reaches the value 1,
around a frequency which is strongly dependent on the cross section and
the mass $m_X$, but not very much on the profile power-law index $\gamma$.
The right part of fig.\ref{freq} shows  the self-absorption coefficent at
the fixed frequency of408MHz as a function of the wimpzilla mass. The
behaviour shown is qualitatively the same for any value of the
cross-section and for different $\gamma$. 

The  coefficent is evaluated for two different values of the cross
section, the first one corresponding to the maximum possible value 
$\sigma v\approx 10^{-28}/m_X^2(\rm GeV)\,\rm cm^3s^{-1},$
(the so-called unitarity bound, see \cite{kamio}) and the
second one corresponding to typical cross sections for supersymmetric
scenarios, $\sigma v=10^{-38}/m_X^2(\rm GeV)\,\rm cm^3s^{-1}$
(see, e.g.,~\cite{bergstrom}).
\begin{figure}[t]
\epsfxsize=27pc 
\epsfbox{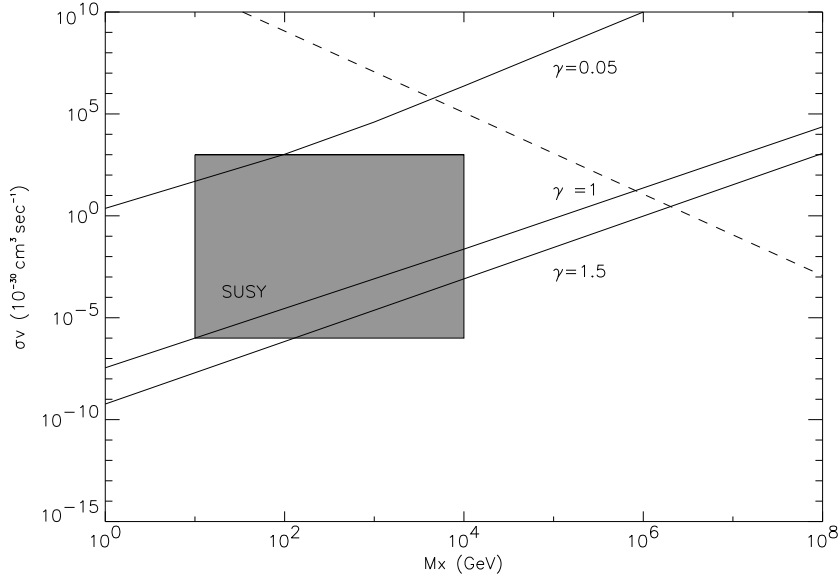}
\caption{Exclusion plot based on the comparison between predicted flux and
radio observations of the galactic centre.The 3 solid curves indicate, for
3 different density profile power law index, the lower edge of the
excluded regions. The dashed line shows, for comparison,
the unitarity bound, $\sigma v\simeq1/m_X^2$.
The shaded region is the portion  of the parameter space occupied
by cosmologically interesting neutralinos (i.e. those leading to
$0.025<\Omega_X h^2<1$; see, e.g. Bergstrom {\em et al} (1997)).
\label{excludo}}
\end{figure}
In figure \ref{predi1} we compare the predicted spectrum with the
observations; we choose a  set of parameters $m_X$, $\gamma$ and $\sigma
v$ in order to reproduce the observed normalisation. It is remarkable that
in this way one can reproduce the observed spectrum over a significant
~range of frequencies.
The submillimeter excess is generally attributed to 
processes in the accretion disk within several Schwarzschild radii
of  the black hole (e.g. \cite {mel}).
The set of dark matter parameters for which the fluxes predicted in our
model are consistent with observations is shown in the exclusion plot of
fig. \ref{excludo}.

\section{Conclusions and perspectives}\label{concludo}

We have  shown that present data on the emission from Sgr A*  are
compatible with  a wide set of dark matter parameters.
The evaluation of synchrotron self-absorption has enabled us to reach an
alternative conclusion to an earlier claim of incompatibility of cuspy
halos with the existence of annihilating wimp dark matter.
We find that the experimental data on the
Sgr A* spectrum at radio wavelengths could be explained by synchrotron
emission of electrons produced in the annihilation of rather massive dark
matter particles, extending up to and beyond TeV masses. This is relevant
to a recent study of coannihilations \cite{boehm}, 
which  suggests that WIMPs with  $\Omega_X h^2=0.2$ can extend up to
several TeV, as well as to very massive  particles (wimpzillas) that are
produced non-thermally in the primordial universe. 
We also find that the synchrotron emission tends to give somewhat sharper
constraints on masses and cross sections than the
observed  gamma-ray fluxes and neutrino limits. This situation could  
change  with more sensitive  gamma-ray 
and neutrino observations anticipated from  forthcoming experiments.

\end{document}